\documentclass[12pt]{iopart}

\usepackage[colorlinks=true,linkcolor=blue,urlcolor=blue,citecolor=blue,pdfusetitle]{hyperref}
\usepackage{bm}% bold math
\usepackage{bbm}
\usepackage{amssymb}
\usepackage{braket}
\usepackage{graphicx}
\usepackage{xcolor}

\newcommand{\SubFig}[2]{\ref{#1}{\color{blue}#2}}

\newcommand{\eqref}[1]{(\ref{#1})}

\newcommand{\Ccal}{\mathcal{C}}

\newcommand{\Scal}{\mathcal{S}}

\newcommand{\1}{\mathbbm{1}}

\newcommand{\Nmath}{\mathbbm{N}}

\usepackage{orcidlink}

\newcommand{\RefAsec}[1]{{\color{black}#1}}
\definecolor{brickred}{RGB}{216, 60, 60}

\newcommand{\RefA}[1]{{\color{black}#1}}
\newcommand{\RefB}[1]{{\color{black}#1}}

\newcommand{\UIMP}{Universidad Internacional Menéndez Pelayo, 28040 Madrid, Spain}
\newcommand{\CSIC}{\href{https://ror.org/009wseg80}{Instituto de Física Fundamental}, \href{https://ror.org/02gfc7t72}{Consejo Superior de Investigaciones Científicas}, Calle Serrano 113b, Madrid 28006, Spain}
\newcommand{\QuARC}{Quantum Advanced Research Center (QuARC), \href{https://ror.org/02gfc7t72}{Consejo Superior de Investigaciones Científicas}, Calle Serrano 113b, 28006 Madrid, Spain}

\begin{document}
	
	\title[Resource-Efficient Digitized Adiabatic Quantum Factorization]{Resource-Efficient Digitized Adiabatic Quantum Factorization}
	
	\vspace{1.5pc}
	
	\author{Felip Pellicer\orcidlink{0009-0002-3686-5586}}
	\address{\UIMP}
	\ead{fpellicer.q@proton.me}

	\author{Juan José García-Ripoll\orcidlink{0000-0001-8993-4624}}
	\address{\QuARC}
	\address{\CSIC}
	\ead{jj.garcia.ripoll@csic.es}
	
	\author{Alan C. Santos$^{\ast}$\orcidlink{0000-0002-6989-7958}}
	\address{\QuARC}
	\address{\CSIC}
	\ead{ac\_santos@iff.csic.es}
	\address{$^{\ast}$Author to whom any correspondence should be addressed}

	\begin{abstract}
		Digitized adiabatic quantum factorization is a hybrid algorithm that exploits the advantage of digitized quantum computers to implement efficient adiabatic algorithms for factorization through gate decompositions of analog evolutions. In this paper, we harness the flexibility of digitized computers to derive a digitized adiabatic algorithm able to reduce the gate-demanding costs of implementing factorization. To this end, we propose a new approach for adiabatic factorization by encoding the solution of the problem in the kernel subspace of the problem Hamiltonian, instead of using ground-state encoding considered in the standard adiabatic factorization proposed by Peng \textit{et al}.~\cite{peng_quantum_2008}. Our encoding enables the design of adiabatic factorization algorithms belonging to the class of \textit{Quadratic} Unconstrained Binary Optimization (QUBO) methods, instead the \RefB{\textit{Polynomial}} Unconstrained Binary Optimization (PUBO) used by standard adiabatic factorization. We illustrate the performance of our QUBO algorithm by implementing the factorization of integers $N$ up to 8 bits. The results demonstrate a substantial improvement over the PUBO formulation, both in terms of reduced circuit complexity and increased fidelity in identifying the correct solution.
		
	\end{abstract}
	
	\section{Introduction}
	\label{sec:introduction}
	
	Digitized Adiabatic Quantum Computing is a promising approach to solve optimization problems by implementing time-continuous adiabatic algorithms with gate-based quantum devices~\cite{Barends:16}. This hybrid model harnesses benefits from adiabatic formulation of quantum algorithms, which allows us to encode the solution of a given problem into the spectrum of a problem Hamiltonian~\cite{Farhi:00,Farhi:01}, and error-correction strategies of gate-based digitized quantum computation~\cite{Shor:96,Barends:14,Terhal:15,Nielsen:Book}. The outcomes of this hybridization are digitized Variational Quantum Algorithms~\cite{McClean_2016,Cerezo:21}---such as \RefB{Quantum Approximate Optimization Algorithm (QAOA)}~\cite{farhi_quantum_2014} and Variational Quantum Eigensolvers~\cite{Peruzzo:14}---have emerged as flexible approaches to incorporate classical processing as crucial ingredient to enhance the efficiency of ditized computers for Noisy Intermediate-Scale Quantum Algorithms~\cite{Zhou:20,Bharti:22,Larkin:22,Zhou:23,Zhang:25}, and quantum simulations~\cite{Lin:22,Ozaeta:22,Tabares:23,Tabares:25,Tabares:26,Hu:25}. 
	
	Of particular interest to this work, digitized adiabatic protocols has been successfully applied to factorization problems~\cite{peng_quantum_2008,anschuetz_variational_2018,hegade_digitized_2021,Hegade:22}. The digitized algorithm proposed in Ref.~\cite{peng_quantum_2008} harnesses the skills of QAOA in tracking the ground state of an adiabatic problem Hamiltonian for factorization belonging to the class of \RefB{\textit{Polynomial}} Unconstrained Binary Optimization (PUBO). As shown in Ref.~\cite{Chermoshentsev:21}, PUBO problems can be reduced to simpler QUBO formulations, which are more efficiently evaluated in the quantum computers. In fact, while QUBO formulations provide Hamiltonians with two-body interaction terms, PUBO requires quantum platforms with native $k$-body interactions~\cite{Chermoshentsev:21}, or the efficient decomposition of these $k$-body terms into two-qubit gates. However, this task involves additional variables and a polynomial reduction that is NP-hard, where even the simplest transformations demand the introduction of additional variables~\cite{Nagies:25}.
	
	In this paper, we introduce a new QUBO formulation of the adiabatic factorization problem that does not require additional variables or NP-hard polynomial reductions. The central idea of our approach is that adiabatic quantum computation does not inherently require the system to be initialized in its ground state. Any energy level can serve as the starting point, provided that the evolution remains adiabatic and transitions to neighboring states are minimized. 
	Building on this insight, we demonstrate that adiabatic factorization can be reformulated as the preparation of the zero-energy eigenstate of a QUBO formula, enabling the development of a novel gate-efficient QAOA algorithm circuit for factorization. Furthermore, by comparing our digitized QUBO-based factorization with the conventional PUBO approach, we show that the proposed model achieves substantially better performance. Also, we analyze the underlying reasons why our QUBO formulation outperforms its PUBO counterpart, which is mainly related to the density of states of the QUBO and PUBO Hamiltonians around the subspace of eigenstates encoding the solution of the adiabatic factorization.
	
	%I also think this needs more precision. There is a gap in jumping from DAQC to QAOA that has not been adequately addressed. Since you change the paradigm, this setence is misleading, because it seems to suggest that QUBO formulation is more advantageous than PUBO in general. Is this true? Also, the last sentence does not really state anything. I don't like paragraphs that merge enunciation of results with a summary of the writing order. This paragraph should be about results. If you want to write about the conclusions, provide here a brief statement of the main ideas and in parenthesis cite the section.
	
	\section{Methods}
	
	\subsection{Adiabatic Quantum Factorization}
	
	The hard instances of factorization involve separating a large integer number $N\in \Nmath_{>0}$ as a product of two odd prime numbers $(q,p) \in \Nmath_{>0}$ satisfying $N = p \times q$. This amounts to finding the zeros of a function
	\begin{equation}
		F(N,p^{\prime},q^{\prime}) = N - \left(2p^{\prime} + 1\right) \times \left(2q^{\prime} + 1\right),\label{Eq:Cost-Function}
	\end{equation}
	where $p=2p'+1$ and $q=2q'+1$.
	As shown by Peng et al.~\cite{peng_quantum_2008}, this problem is solved by a quantum computer that prepares the ground state $\ket{\Psi_\mathrm{sol}}$ of the problem Hamiltonian $\hat{H}_{\mathrm{QP}} = F(N,\hat{p}^\prime,\hat{q}^\prime)^2$ defined using the observables $\hat{p} = \sum_{i=1}^{n_p} 2^{i-1} \hat{x}_i$ and $\hat{q} = \sum_{j=1}^{n_q} 2^{j-1} \hat{y}_j$ that reconstruct natural numbers from independent sets of qubits $\hat{x}_i,\hat{y}_j\in\{0,1\}$ in a quantum register. In this formulation, the problem Hamiltonian is the polynomial
	\begin{equation}
		\hat{H}_\mathrm{QP} = \bigg[ N \1 - \bigg( \sum_{\ell=1}^{n_p} 2^\ell \hat{x}_\ell + \1 \bigg)
		\bigg( \sum_{m=1}^{n_q} 2^m \hat{y}_m + \1 \bigg) \bigg]^2\,,
		\label{eq:quadratic_problem_hamiltonian}
	\end{equation}
	and the ground state is the set of bits that minimize the energy
	\begin{equation}
		\ket{\Psi_{\mathrm{sol}}} = \ket{p^{\prime}}\otimes \ket{q^{\prime}} = \left(\ket{x_{1}}\otimes \cdots \otimes \ket{x_{n_p}}\right) \otimes \left(\ket{y_1}\otimes \cdots \otimes \ket{y_{n_q}}\right). \label{Eq:Psi_Sol}
	\end{equation}
	The largest number of qubits required to encode this problem are~\cite{peng_quantum_2008}
	\begin{equation}
		n_p = \Scal\,\big(\lfloor \sqrt{N} \rfloor_O\big) - 1 ,\quad 
		n_q = \Scal\bigg(\left\lfloor \frac{N}{3} \right\rfloor \bigg) - 1 ,
		\label{eq:factors_num_bits}
	\end{equation}
	where $\lfloor X \rfloor$ denotes the \textit{largest integer} not larger than $X$, $\lfloor X \rfloor_O$ is the \textit{largest odd integer} not larger than $X$, and $\Scal(X)\sim\lceil{\log_2(X)}\rceil$ is the \textit{smallest number} of bits required for representing $X$.
	
	The formulation of factoring as a ground state search means that this problem can be solved using adiabatic quantum computers, quantum annealers and any other adiabatic state preparation methods for digital quantum computers. Unfortunately, the Hamiltonian in Eq.~\eqref{eq:quadratic_problem_hamiltonian} corresponds to a family of PUBO problems that involve non-local three- and four-qubit interactions $Z_{3} = \hat{\sigma}_\ell^z \hat{\sigma}_m^z \hat{\sigma}_k^z$ and $Z_{4} = \hat{\sigma}_\ell^z \hat{\sigma}_m^z \hat{\sigma}_k^z \hat{\sigma}_n^z$, gathering qubits from arbitrary locations in the quantum register. These non-local terms complicate the use of quantum annealers or other quantum optimizers, because they require establishing interactions among distant quantum objects---which may be impossible, depending on the architecture, or may require the use of multiple auxiliary qubits~\cite{Babbush:13}.
	
	In digital applications, these non-local terms complicate the implementation of evolution with the Hamiltonian, $\exp(-i\hat{H}_{\mathrm{QP}}\delta t)$, which requires two-, three- and four-partite controlled phases, $\exp(-iZ_3)$ or $\exp(-iZ_4)$, which must be decomposed into multiple applications of native two-qubit gates. For these reasons, PUBO-based digitized factorization has been limited to efficient (high fidelity) factorization achieved for numbers up to $91$~\cite{hegade_digitized_2021,Agresti:25}. For higher numbers, the success probability for factorization drastically decreases below $20\%$~\cite{Lin:22}, even using machine learning techniques.
	
	\subsection{Null-subspace encoding and the QUBO Hamiltonian}
	
	Let us now discuss an alternative formulation of the factoring problem that addresses all the challenges discussed above. The key idea in our proposal is the realization that the adiabatic theorem equally applies to the preparation of ground state as to the creation of higher excited states---a fact demonstrated experimentally, for instance, in Ref.~\cite{Hu:25}. With this in mind, let us define a new factorization protocol as the preparation of the zero eigenstates $\hat{H}_\mathrm{LP}\ket{\Psi_{\mathrm{sol}}}=0$ of a Hamiltonian defined as
	\begin{equation}
		\hat{H}_\mathrm{LP} := F(N,\hat{p}^{\prime},\hat{q}^{\prime}) = N \1 - \bigg( \sum_{\ell=1}^{n_p} 2^\ell \hat{x}_\ell + \1 \bigg)
		\bigg( \sum_{m=1}^{n_q} 2^m \hat{y}_m + \1 \bigg) ,
		\label{eq:linear_problem_hamiltonian}
	\end{equation}
	
	This is a \textit{linearized} alternative to the $\hat{H}_\mathrm{QP}$ introduced in Eq.~\eqref{eq:quadratic_problem_hamiltonian}. While the co-prime factors are defined by the same quantum eigenstates of both models, with $\bra{\Psi_{\mathrm{sol}}}\hat{H}_\mathrm{LP}\ket{\Psi_{\mathrm{sol}}}=\bra{\Psi_{\mathrm{sol}}}\hat{H}_\mathrm{QP}\ket{\Psi_{\mathrm{sol}}}=0$, in $\hat{H}_\mathrm{LP}$ the solution is an excited state living in the  \textit{kernel subspace} of $\hat{H}_\mathrm{LP}$. Most important, the problem Hamiltonian $\hat{H}_\mathrm{LP}$ is now a QUBO model that only contains two-body interactions.
	
	%Finally, and equally important, the eigenenergies of $\hat{H}_\mathrm{LP}$ in the computational basis are given by the values of the quadratic function $F(N,p^{\prime},q^{\prime})$ over all integer values of $p^{\prime}$ and $q^\prime$. This spectrum grows more slowly, a fact that affects (positively) the instantaneous energy gaps and the performance of the adiabatic preparation protocols. 
	
	%\RefA{The linearization strategy presented here is generally applicable whenever the standard adiabatic problem Hamiltonian can be written as $\hat{H}_\mathrm{P} = \hat{H}_0^{2n}$ for some Hamiltonian $\hat{H}_0$ and integer $n \geq 1$, with the solution contained in its kernel, $\hat{H}_0 \ket{\Psi_{\mathrm{sol}}} = 0$. The simplified Hamiltonian is QUBO only if $\hat{H}_0$ is QUBO, what is true for the factorization protocol considered here. However, if we consider extending factorization to three integers, $N=pqr$, we propose to encode the solution in the kernel of the new problem Hamiltonian $\hat{H}^{pqr}_0 = N \1 - \hat{p}\hat{q}\hat{r}$. In this case, the squared problem Hamiltonian $\hat{H}^{pqr}_\mathrm{P} = (\hat{H}^{pqr}_0)^2$ contains up to six-body $Z$ interactions, which can be reduced to three-body terms with our approach. Despite this substantial simplification, $\hat{H}^{pqr}_0$ is not QUBO, as it is now trilinear across three registers $(p,q,r)$.}
	
	This simplified formulation is challenging for quantum computers and quantum simulators because it involves long-range interactions, and because the preparation of an excited state lifts all the protections against decoherence expected from an adiabatic quantum computing framework~\cite{albash_adiabatic_2018}. \RefA{Moreover, encoding the solution in an excited state introduces additional challenges, since the gap protection available for the ground state cannot be guaranteed, and the adiabatic gap condition must be satisfied with respect to both higher- and lower-energy states. Nevertheless, as we show below, these limitations can be circumvented by employing the linearized Hamiltonian as the \textit{ansatz} Hamiltonian in a QAOA circuit, while defining the variational cost function in terms of the modified Hamiltonian $\hat{H}_{\mathrm{LP}}^{\mathrm{abs}}=|\hat{H}_{\mathrm{LP}}|$. In this way, the target eigenstate of $\hat{H}_{\mathrm{LP}}$ becomes the ground state of the cost Hamiltonian, allowing the optimization to be formulated as a ground-state search.} The resulting model is still much simpler than the PUBO formulation and opens the door to efficient digitized approximations. In particular, this Hamiltonian is of comparable difficulty to other models used in variational quantum computation~\cite{Harrigan:21}, and for that reason the following sections will explore variational QAOA protocols based on this new model.

	\subsection{Variational Algorithm for Adiabatic Factorization}
	
	Similarly to digitized adiabatic dynamics, variational algorithms are an efficient paradigm to simulate quantum annealing using gate-based quantum devices. Among the class of \RefB{variational quantum algorithms}, a way to connect quantum annealers and digital computers is provided by the QAOA~\cite{farhi_quantum_2014,Zhou:20}. The algorithm aims to approximate the ground state of a cost Hamiltonian whose minimum encodes the optimal solution. At its core, QAOA applies the same strategy as AQC, driving the system from an easy-to-prepare ground state $\ket{\psi_{\mathrm{inp}}}$ of the initial Hamiltonian $\hat{H}_{\mathrm{ini}}$, to the output solution state $\ket{\psi_{\mathrm{sol}}}$ encoded in the \textit{ground state} of the problem Hamiltonian $\hat{H}_{\mathrm{P}}$. To this end, QAOA constructs a parametrized quantum state (ansatz) by sequentially applying two alternating types of unitaries derived from these two main Hamiltonians. Using the QAOA terminology, the problem Hamiltonian $\hat{H}_{\mathrm{P}}$ is identified as the cost Hamiltonian, which encodes the objective function of the optimization problem. On the other hand, $\hat{H}_{\mathrm{ini}}$ is named mixer Hamiltonian $\hat{H}_\mathrm{M}$, which introduces transitions between computational basis states to enable exploration of the solution space. 
	
	In the particular case of \RefB{adiabatic quantum factorization}, the QAOA algorithm starts with the initial state $\ket{\psi_0} = \ket{+}^{\otimes n}$, $\ket{\pm} = (\ket{0} \pm \ket{1})/\sqrt{2}$, as it is a uniform superposition of all computational states of $n$ bits. For this reason, the mixer Hamiltonian is defined as $\hat H_\mathrm{M} = \Omega \sum_{k=1}^{n} \hat{\sigma}^{x}_{n}$, guaranteeing that $\ket{\psi_0}$ is the ground state of the standard adiabatic protocol. On the other hand, in order to keep the same mixer Hamiltonian, the QUBO approach requires a different state initialization, as we need to start the evolution from the null-subspace of $\hat{H}_{\mathrm{LP}}$. This is done by choosing $\ket{\psi^{\mathrm{LP}}_0} = \ket{+}\ket{-}\cdots\ket{+}\ket{-}$. Similar to $\ket{\psi_0}$, $\ket{\psi^{\mathrm{LP}}_0}$ also represents a uniform superposition of all computational states. The only concern we need to take into account is the case in which the number of qubits $n$ is not odd, as in this case the state $\ket{\psi^{\mathrm{LP}}_0}$ is not the zero energy eigenstate of $\hat{H}_{\mathrm{LP}}$. However, our QAOA algorithm is robust against this small deviation from the null-subspace of the QUBO Hamiltonian. Indeed, our QAOA algorithm performs efficiently even if we start both algorithms in state $\ket{\Psi_0}$~\cite{repo}, which is a disadvantageous scenario for $\hat{H}_{\mathrm{LP}}$.
	
	For the integration with classical optimizers, we employ cost functions defined in terms of the Hamiltonian that serves as the reference for constructing the QAOA circuit. Specifically, the PUBO-based circuit minimizes the cost function 
	$
	\mathcal{C}_{\mathrm{QP}} = \langle \hat{H}_{\mathrm{QP}} \rangle
	$,
	whereas the QUBO-based approach relies on
	\RefA{$
		\mathcal{C}_{\mathrm{LP}} = \langle \hat{H}_{\mathrm{LP}}^{\mathrm{abs}} \rangle
		$. Note that our cost function is $\langle \hat{H}_{\mathrm{LP}}^{\mathrm{abs}} \rangle$, which is different from $\vert\langle  \hat{H}_{\mathrm{LP}}  \rangle\vert$. Evaluating this function does not introduce any additional computational complexity to the protocol. In fact, because $\hat{H}_{\mathrm{LP}}$ and $\hat{H}_{\mathrm{LP}}^{\mathrm{abs}}$ are diagonal in the computational basis, $\{\ket{n}\}$, we write $\hat{H}_{\mathrm{LP}} = \sum_n E_n \ket{n}\bra{n}$ and $\hat{H}_{\mathrm{LP}}^{\mathrm{abs}}= \sum_n |E_n| \ket{n}\bra{n}$. This shows that no additional complex spectral decomposition is required to evaluate $\mathcal{C}_{\mathrm{LP}}$, which shows that $\mathcal{C}_{\mathrm{LP}}$ and $|\langle \hat{H}_{\mathrm{LP}}\rangle|$ can be evaluated with identical computational complexity.} Therefore, the optimization landscape of $\mathcal{C}_{\mathrm{LP}}$ is explored by the same classical algorithm employed for $\mathcal{C}_{\mathrm{QP}}$, ensuring consistent performance evaluation across both methods. We also consider other strategies to define the cost function, which provides similar results~\cite{repo}, \RefA{but the best convergence to the right solution was obtained using $\langle \hat{H}_{\mathrm{LP}}^{\mathrm{abs}} \rangle$}.

	\section{Results and Discussions}
	
	In this section we present the main results of this work. 
	%\RefA{As we shall see, while the proposed method considered here might require additional spectral analysis when dealing with arbitrary Hamiltonians---in order to keep a finite gab between the null-subspace and its neighboring energy levels---our results demonstrate that our method can be helpful in practical scenarios. The efficiency of our approach is mainly supported by two results. First, QAOA can be understood as a Trotterization with the QAOA circuit parameters serving as the discrete counterparts of the adiabatic Hamiltonian time-dependent~\cite{Pablo:26}. Second, the validity condition for high-fidelity digitized quantum annealing (Trotterization) is not restrict to ground-states digitized evolutions~\cite{Santos:25}. Therefore, by a suitable choice of the \textit{cost function} and \textit{problem Hamiltonian}, we show now that QAOA can be used in tasks beyond ground-state-based protocols.}
	We start by analyzing the performance of the PUBO and QUBO approaches for factorization of a 5-bits number. Then, we increase the complexity of the problem, factorizing up to 8-bits numbers. To end, we discuss about the mechanism behind the enhancement promoted by the QUBO Hamiltonian, leading to an intuitive explanation of why our proposal outperforms the PUBO Adiabatic QAOA model. \RefA{See~\ref{Appendix:SimulationSettings} for further details on the QAOA numerical implementation}.
	
	\subsection{Factorizing the number 25}
	
\begin{figure}[t!]
		\centering
		\includegraphics[width=1.0\linewidth]{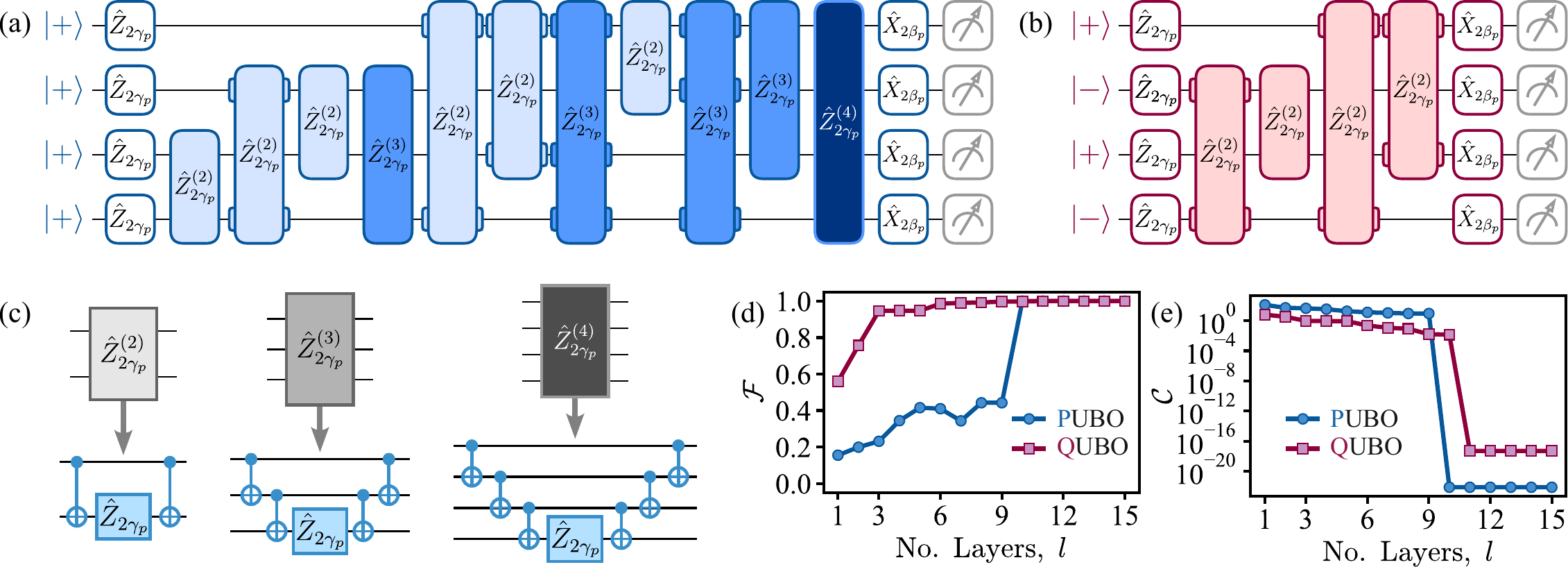}
		\caption{Single layer circuit of the (a) PUBO and (b) QUBO algorithms for factoring $N = 25$. While PUBO circuit requires three- and four-qubit gates, which can be decomposed according the circuits in (c), the QUBO algorithm requires two-qubit gates only. We define the circuit gates as $\hat{Z}_{2\gamma_p} = \exp( i \gamma_p \hat{\sigma}^{z}_k )$,
			$\hat{X}_{2\beta_p} = \exp( i \beta_p \hat{\sigma}^{x}_k )$,	
			$\hat{Z}_{2\gamma_p}^{(2)} = \exp( i \gamma_p \hat{\sigma}^{z}_k \hat{\sigma}^{z}_m )$,		
			$\hat{Z}_{2\gamma_p}^{(3)} = \exp( i \gamma_p \hat{\sigma}^{z}_k \hat{\sigma}^{z}_m \hat{\sigma}^{z}_n )$,		
			$\hat{Z}_{2\gamma_p}^{(4)} = \exp( i \gamma_p \hat{\sigma}^{z}_k \hat{\sigma}^{z}_m \hat{\sigma}^{z}_n \sigma^{z}_l )$. (d) Fidelity of finding the right solution as function o the number of layers for the PUBO and QUBO algorithms, followed by (e) the behavior of the cost function for each model.}
		\label{Fig:5-Digit}
	\end{figure}

	In order to demonstrate the enhancement promoted by our protocol, first let us to discuss its performance by factorizing small numbers. In particular, we consider the factorization of $N=25$, which is a 5-bits number. To this, the adiabatic algorithm requires a minimum of 4 qubits, and the QAOA circuit is shown in Figs.~\SubFig{Fig:5-Digit}{a} and~\SubFig{Fig:5-Digit}{b}. The set of information in Fig.~\ref{Fig:5-Digit} is the first result of our work.
	
	From Figs.~\SubFig{Fig:5-Digit}{a} and~\SubFig{Fig:5-Digit}{b} it is possible to identify the significant simplification of the QAOA circuit to implement the QUBO adiabatic Hamiltonian for the digitized algorithm, in comparison with the standard protocol. This enhancement become even \RefB{stronger} if we take into account that three-qubit and four-qubit operations in Figs.~\SubFig{Fig:5-Digit}{a} need to be decomposed into two-qubit gates to be compared to Fig.~\SubFig{Fig:5-Digit}{b}. As shown in Fig.~\SubFig{Fig:5-Digit}{c}, taking as reference the decomposition of $ZZ$-interaction into CNOTs and local $Z$ rotations, a $n$-body $ZZ$ interaction needs $2n$ CNOTs. 
	
	This analysis shows that the 34-CNOTs circuit, per layer, required by the standard factorization is replaced by a 8-CNOTs circuit per layer, if we use the linear protocol. In addition, even most relevant, as shown in Figs.~\SubFig{Fig:5-Digit}{d} and~\SubFig{Fig:5-Digit}{e}, the QUBO algorithm outperform the standard approach. While fidelity measurements should be avoided in QAOA algorithms, we also show the fidelity in this case as an additional metric of the performance of the circuits. But, the relevant result for QAOA is shown in ~\SubFig{Fig:5-Digit}{e}, where we present the cost function behavior in function of the number of layers.

	\subsection{Factorizing higher numbers}
	
\begin{figure}[t!]
		\centering
		\includegraphics[width=\linewidth]{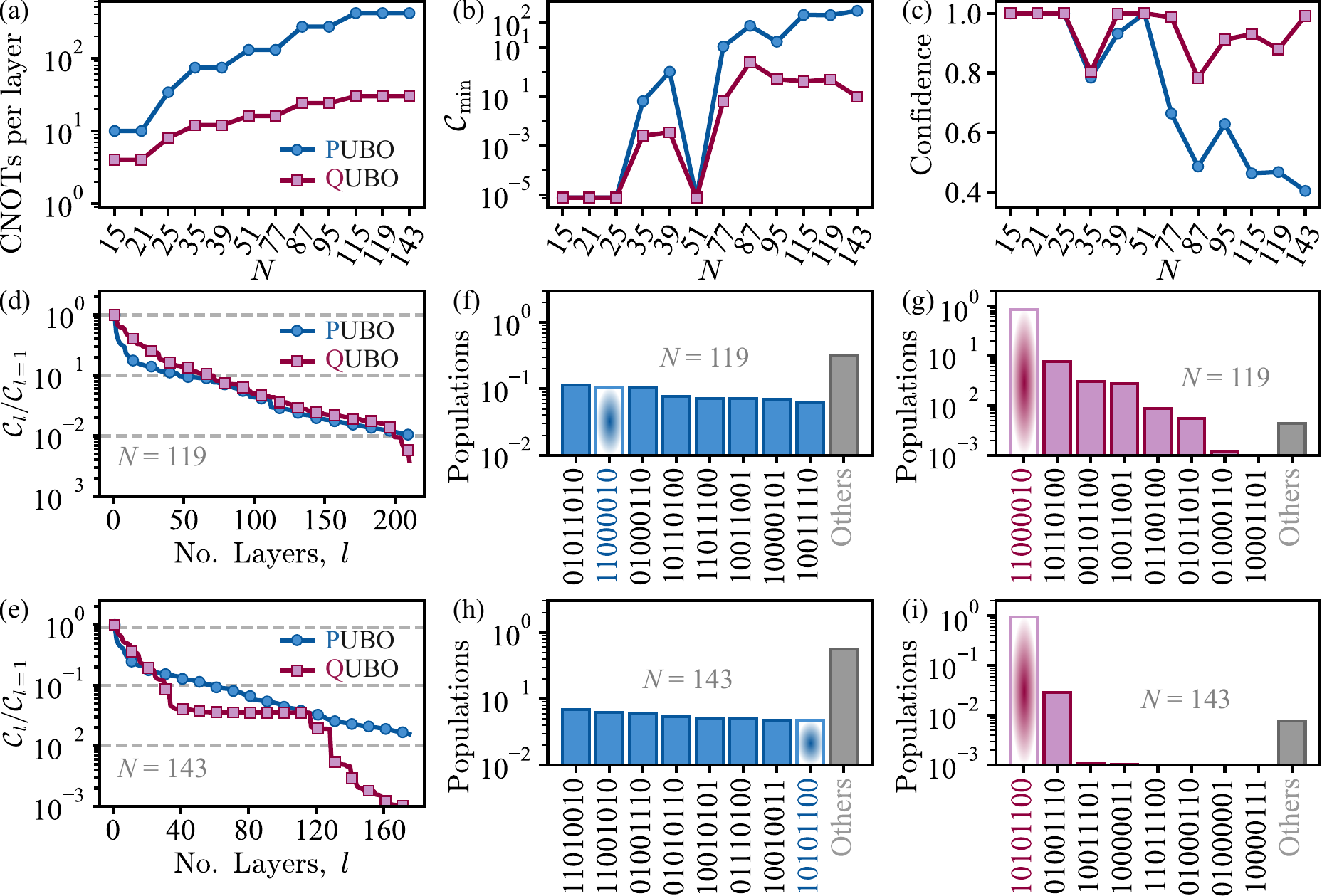}
		\caption{(a) Single-layer circuit scaling of PUBO and QUBO algorithms in terms of the number of CNOTs per layer as a function of $N$. In (b) and (c), respectively, we show the behavior of the smallest (final) value of the cost function $\Ccal_\mathrm{min}$ and the confidence of each algorithm as function of $N$. For the number $N=119$ we present (d) the normalized cost $\Ccal_{l}/\Ccal_{l=1}$ as function $l$, showing the behavior of the cost for each algorithm. Complementarily, we show in (f) and (g) the population distribution in the computational basis ordered from highest to lowest populated states for the PUBO and QUBO algorithms, respectively, where we highlight the corresponding desired solution. Similar data are shown for the number $N=143$ in (e,h,i). ``Others" constitutes the sum of populations of all computation states not displayed.}
		\label{Fig:Scaling}
	\end{figure}

	Going beyond the few digit numbers factorization we use the standard and QUBO approach to factorize all numbers shown in~\ref{Appendix:Table}, Table~\ref{Table_factor}. For all cases, we repeat the procedure aforementioned using the cost function as main quantity to guide the classical optimizers of our QAOA protocols. To provide a robust characterization of the performance of each protocol, we first investigate the number of CNOTs per layer, in Fig.~\SubFig{Fig:Scaling}{a}, and the (absolute) lowest values achieved for the cost function $\Ccal_{\mathrm{min}}$ for different factorized numbers $N$, shown in Fig.~\SubFig{Fig:Scaling}{b}. 
	
	As the main conclusion, the scaling advantage of the \textit{QUBO factorization} over the standard formulation is clearly demonstrated---at least, to the cases considered here. The presence of three- and four-body interaction terms in the PUBO Hamiltonian introduces a noticeable degradation in the performance of the PUBO-QAOA, while the QUBO formulation requires substantially fewer CNOT gates per layer. Consequently, even in cases where both approaches efficiently identify the correct factors (for instance, when $N \leq 77$), the QUBO digitized factorization achieves this with a significantly lower gate count per layer. Moreover, the most remarkable outcome of this work arises when considering larger numbers such as $N = 119$ and $143$, each requiring eight qubits. In these cases, the QUBO algorithm exhibits a clear advantage in efficiency and reliability for identifying the correct solution, outperforming the standard PUBO formulation for adiabatic factorization.

	A key feature of adiabatic quantum algorithms lies in their ability to amplify the probability amplitudes of states corresponding to valid or near-valid solutions, such as search-based algorithms. The overall success of the algorithm therefore depends on its capacity to populate these promising states with high probability. For this reason, an important performance metric is the \textit{confidence} of the output state---that is, the probability weight associated with the correct or near-correct solutions---. To quantify this confidence, we measure the overall uncertainty of the distribution using \textit{Shannon entropy}, $H = -\sum_i p_i \log(p_i)$, given the normalized amplitudes (in our case populations) $p_i$ of all states. Because a low entropy value indicates that the distribution is sharply peaked, it suggests that one outcome strongly dominates and thus the prediction is highly confident. Conversely, a high entropy implies that several outcomes have comparable amplitudes, which corresponds to a higher level of uncertainty. To obtain a normalized confidence score, one can define
	\begin{equation}
		C = 1 - \frac{H}{H_{\mathrm{max}}},
	\end{equation}
	where $H_{\mathrm{max}}$ is the entropy of a uniform distribution over all possible outcomes. In other words, we compute it using the input state, or we can use directly that in this case we have $p_i = 1/2^n$, where $n$ is the number of qubits in the system. This yields a confidence measure $C \in [0, 1]$, where $C = 1$ corresponds to maximum confidence and $C = 0$ to complete uncertainty.

	This significant enhancement in the confidence is confirmed by the monotonically decreasing behavior of the cost function, and the distribution of probabilities of the output state from each protocol, as shown in Fig.~\ref{Fig:Scaling}. As we can see from Figs.~\SubFig{Fig:Scaling}{d} and~\SubFig{Fig:Scaling}{e}, the \textit{normalized} cost functions for factorization of the highest factorized numbers, namely $119$ and $143$, behave as expected during the QAOA implementation. By normalizing the cost function by the single layer cost function $\Ccal_{l=1}$, we observe that the QAOA circuit for the QUBO Hamiltonian can reduce the initial energy by up to 3 orders of magnitude (for $N=143$), while the standard model is able to reduce the cost function up to 2 order of magnitude. \RefA{In particular, the factorization of $N=143$ exhibits an extended plateau in which the QUBO cost decreases only marginally over a wide range of layers before dropping further. Even in this less favorable scenario, the QUBO formulation still outperforms its PUBO counterpart.}
	
	Complementarily, we identify the solution prime numbers, $q^{\prime}$ and $p^{\prime}$, by reading the final populations distributions in the computational bases in Figs.~\SubFig{Fig:Scaling}{f}-\SubFig{Fig:Scaling}{i}, sorted from the most populated states, with the accumulated population ``Others" representing all other states. As we can see, the good confidence of the QUBO approach is most due to high population transfer to given state, that can be tested as solution of the problem. As indicated by the confidence parameter, in Fig.~\SubFig{Fig:Scaling}{c}, the distribution of probabilities for both number is flatted, leading to a uncertainty of the right solution. Most drastically, by testing all states we can identify the solution and, in both cases, they are not placed as most populated states.
	
	%\RefA{From Fig.~\ref{Fig:Scaling}, we observe that the QUBO factorization protocol for the number $N=35$ and $N=143$ provides the right answer with high confidence (above $80\%$). These two cases are particularly important because they are associated to problem Hamiltonians with degenerate target subspace. In fact, as detailed in~\ref{Appendix:Table}, the states $\{\ket{01110},\ket{11010}\}$ are eigenstates of $\hat{H}_{\mathrm{LP}}$ and $\hat{H}_{\mathrm{QP}}$ with null eigenvalue for $N=35$, and similarly $\{\ket{10101100}, \ket{01110100}\}$ are degenerate states with null eigenvalue for the case $N=143$. The advantage in factorization is mainly due to the fact that the, in all cases, the kernel subspace is only composed by solutions of the problem.}
	
	\subsection{Hamiltonian spectral density analysis}
	
	To explain the enhanced performance of the QUBO algorithm, we conjecture that QAOA is mainly impacted by (i) the convergence of optimization---due to local minima trapping~\cite{Cerezo:21,Lyngfelt:25} or barren plateaus~\cite{McClean:18,Arrasmith:22}, for instance---and (ii) probability of escape from the ideal solution and adiabatic passage~\cite{sarandy_consistency_2004,Amin:09}---as digitized adiabatic algorithms are also affected by the adiabatic minimum gap condition~\cite{Santos:25}. Regarding (ii), we can state that for a fixed minimum gap, the higher the density of accessible states, the higher the probability that the quasi-adiabatic QAOA evolution drives the system away from the target computational subspace. For this reason, we investigate the spectral decomposition of the problem QUBO and PUBO Hamiltonians, where we observe that the density of states of the PUBO Hamiltonian, around the solution subspace, is much higher than its QUBO counterpart.

	\begin{figure}[t!]
		\centering
		\includegraphics[width=\linewidth]{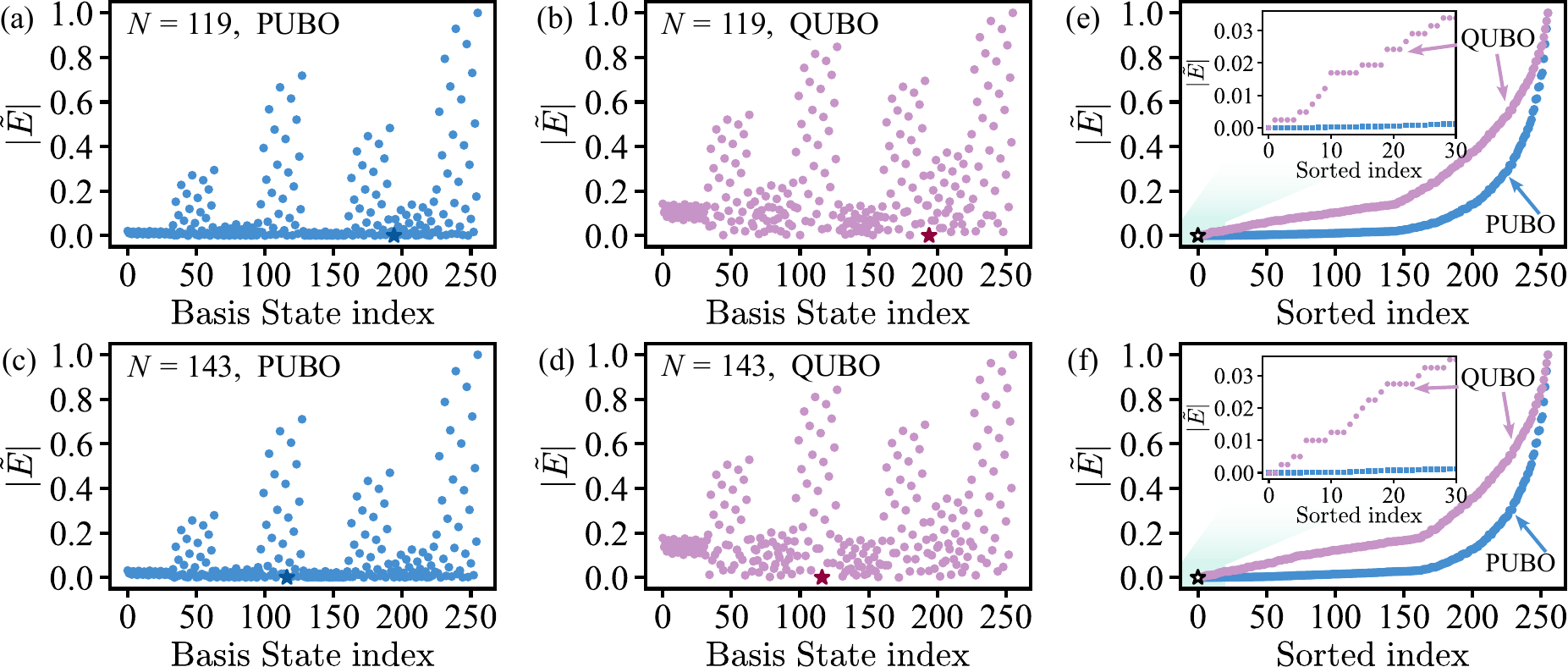}
		\caption{Normalized energy spectra $|\tilde{E}_{n}| = |E_{n}|/E_{\mathrm{max}}$ of the (a) PUBO Hamiltonian and (b) QUBO Hamiltonian used for factorization of $N=119$, where the $x$-axis shows the decimal encoding of computational basis states. The eigenstate corresponding to the solution ($\widetilde{E}=0$) is marked with red star symbol. Similar plots are shown for the number $N=143$ in (c) and (d). The panels (e) and (f) show the same spectra as (a,b) and (c,d), respectively, but we use a basis ordering based on the energy $|\tilde{E}_{n}|$. This graphs, including the inset plots, allow us to observe how the minimum norm $|\tilde{E}_{n}|$ search is challenged by the high density of states in the PUBO algorithm compared to the QUBO approach.}
		\label{Fig:Scatter}
	\end{figure}
	
	In Fig.~\ref{Fig:Scatter} we show the scatter plot of the normalized eigenenergies $\tilde{E}_{n} = E_{n}/E_{{\max}}$, where $E_{{\max}}$ is the highest eigenenergy, of the problem Hamiltonians as function of the basis state index. From the results shown in Figs.~\SubFig{Fig:Scatter}{a--d}, it is possible to understand the reason why the QUBO Hamiltonian provides enhanced results in comparison with its PUBO counterpart. It is evident the high density of states with energy near to the ground state of the PUBO Hamiltonian, while the QUBO spectrum is much less dense around the eigenenergy $\tilde{E}=0$.
	
	The impact of high density spectrum of the PUBO Hamiltonian becomes even more evident when we consider the ordering of the eigenenergies, as shown in Figs.~\SubFig{Fig:Scatter}{e} and~\SubFig{Fig:Scatter}{f}, and their inset plots. We find that the spectrum of $\hat{H}_\mathrm{LP}$ is more widely spread around the solution, leading to larger separations between relevant energy levels near to the kernel subspace. \RefA{However, $\hat{H}_\mathrm{LP}$ presents degeneracy for excited states (see inset in Figs.~\SubFig{Fig:Scatter}{e} and~\SubFig{Fig:Scatter}{f}), which may contribute to the extended plateau in the cost observed for $N=143$ in Fig.~\SubFig{Fig:Scaling}{e}}. On the other hand, the flatted spectrum of the PUBO Hamiltonian around low-energy states also extends to higher energy states. This constitutes a relevant barrier to classical optimizers in supporting an efficient redistribution of amplitude during the variational evolution. Eventually, this may be the main reason behind the population distribution profile shown in Figs~\SubFig{Fig:Scaling}{f} and~\SubFig{Fig:Scaling}{h}.
	
	\begin{figure}[t!]
		\centering
		\includegraphics[width=\linewidth]{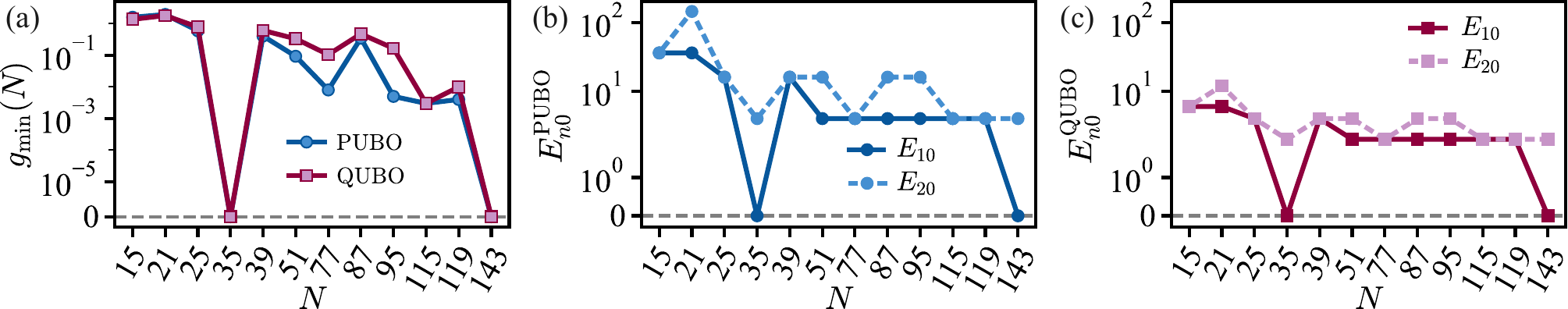}
		\caption{\RefAsec{(a) Minimum gap between the ground state subspace and its closed neighboring energy level. Energy gap $E_{n0}$ between the three lowest energetic states of the problem Hamiltonian used as cost function of the (b) PUBO and (c) QUBO factorization protocols.} }
		\label{Fig:Gaps}
	\end{figure}
	
	\RefAsec{This interpretation is further supported by analyzing the minimum energy gap and the energy separation associated with the degeneracy of the three lowest-lying states. In Fig.~\SubFig{Fig:Gaps}{a}, we show the minimum gap along the time-dependent adiabatic protocol governed by the Hamiltonian $\hat{H}(s) = (s-1) \hat{H}_{\mathrm{M}} + s \hat{H}_{\mathrm{fin}}$, where $\hat{H}_{\mathrm{fin}} = \hat{H}_{\mathrm{LP}}$ and $\hat{H}_{\mathrm{fin}} = \hat{H}_{\mathrm{QP}}$ correspond to the QUBO and PUBO protocols, respectively, and $s \in [0,1]$ is a dimensionless temporal parameter. We define the minimum gap as $g_{\mathrm{min}}(N) = \min_{n>0} \left[\min_{s} \left(E_{n}(s)-E_{0}(s)\right)\right]$. For both models, except in the degenerate cases $N=35$ and $N=143$, the minimum gap of the Hamiltonians remains finite for all values of $N$ considered in this work, with the largest gap for the QUBO model.
		
		The minimization over the normalized time parameter $s$ is performed over the open interval $s \in [0,1]$, including also the endpoints where degeneracies may occur. As discussed above and illustrated in Figs.~\SubFig{Fig:Gaps}{b} and~\SubFig{Fig:Gaps}{c}, by computing the energy difference $E_{n0} = E_{n} - E_{0}$ between the ground state energy $E_{0}$, and the energy levels $n=1$ and $n=2$, we see that the ground state becomes degenerate for $N=35$ and $N=143$ at $s=1$, as $\hat{H}(s=1) = \hat{H}_{\mathrm{fin}}$. Thus, for the cases considered here, the apparent gap closing is associated with the degeneracy of the final Hamiltonian rather than with a gap closing during the adiabatic evolution.}
	
	\section{Conclusion}
	
	In this paper, we introduced an efficient digitized algorithm for implementing adiabatic factorization, with enhanced success fidelity and circuit complexity. The first major outcome of our study is that the circuit-level enhancement achieved through our approach holds pivotal importance for experimental implementations of digitized quantum computing, as it leads to a substantial reduction in the total algorithmic runtime. Moreover, across all cases analyzed, our QAOA implementation based on the QUBO Hamiltonian consistently yields the correct factorization result, characterized by a highly populated target state that ensures strong confidence in the outcome. \RefB{In contrast, the same strategy fails to achieve comparable performance when applied to the PUBO Hamiltonian formulation of the factorization problem. Consequently, our results provide new insights into the ongoing debate regarding the relative merits of PUBO and QUBO encodings in QAOA-based optimization. Contrary to previous studies~\cite{Stein:23}, which reported that PUBO formulations generally deliver superior performance while requiring fewer qubits, we demonstrate that an appropriate choice of the problem Hamiltonian and cost function can enable the QUBO formulation of factorization to outperform its PUBO counterpart.}
	
	The most relevant difference between our QUBO proposal for adiabatic factorization and its PUBO formulation is the encoding of the solution in a state in the middle of the Hamiltonian spectrum, rather than in the ground state. While this represents a challenge for quantum annealers---since the ground-state protection against decoherence cannot be exploited~\cite{Albash:2012,Albash:15}---it does not pose a problem for digitized quantum annealers, where decoherence is instead governed by local single-qubit decay. In this scenario, the QUBO approach is expected to be more robust than the PUBO one, as QUBO requires significantly fewer gates per block, thereby reducing the total runtime of the algorithm. Moreover, this reduced gate demand is also expected to constitute an advantage of QUBO-based factorization over its PUBO counterpart in any realistic setting with finite gate error rates.
	
	Looking ahead, our results motivate further exploration of digitized adiabatic protocols in broader optimization and number-theoretic settings. In particular, extending our framework to larger composite numbers and more complex problem Hamiltonians may offer deeper insights into the practical advantages of QUBO-based implementations. Moreover, studying the interplay between circuit depth, noise resilience, and Hamiltonian structure could guide the development of more resource-efficient quantum algorithms tailored for near-term devices.
	
	\section*{Data availability statement}
	
	The data and codes that support the findings of this study will be openly available in Zenodo following an embargo at the following URL: \href{https://zenodo.org/records/18414560}{https://zenodo.org/records/18414560}.
	
	\section*{Conflict of interest}
	
	The authors declare no competing interests.
	
	\section*{Funding}
	
	JJGR acknowledges support from the EuroHPC Joint Undertaking under project QEC4QEA (Ref. 101194322), and by project PCI2025-163266 funded by MICIU/AEI/10.13039/501100011033. ACS is supported by the Comunidad de Madrid through the program Ayudas de Atracción de Talento Investigador ``César Nombela", under Grant No. 2024-T1/COM-31530 (Project SWiQL). 
	
	\section*{Author contributions}
	
	All authors have contributed equally to this work.
	
	\section*{ORCID iDs} 
	
	Felip Pellicer\orcidlink{0009-0002-3686-5586} \href{https://orcid.org/0009-0002-3686-5586}{https://orcid.org/0009-0002-3686-5586}
	
	\noindent Juan José García-Ripoll\orcidlink{0000-0001-8993-4624} \href{https://orcid.org/0000-0001-8993-4624}{https://orcid.org/0000-0001-8993-4624}
	
	\noindent Alan C. Santos\orcidlink{0000-0002-6989-7958} \href{https://orcid.org/0000-0002-6989-7958}{https://orcid.org/0000-0002-6989-7958} 
	
	\section*{References} 
	
	%\bibliography{references.bib}

\begin{thebibliography}{10}
		\expandafter\ifx\csname url\endcsname\relax
		\def\url#1{{\tt #1}}\fi
		\expandafter\ifx\csname urlprefix\endcsname\relax\def\urlprefix{URL }\fi
		\providecommand{\eprint}[2][]{\url{#2}}
		
		\bibitem{peng_quantum_2008}
		Peng X, Liao Z, Xu N, Qin G, Zhou X, Suter D and Du J 2008 {\em Physical Review
			Letters\/} {\bf 101} 220405
		
		\bibitem{Barends:16}
		Barends R, Shabani A, Lamata L, Kelly J, Mezzacapo A, Heras U~L, Babbush R,
		Fowler A~G, Campbell B, Chen Y, Chen Z, Chiaro B, Dunsworth A, Jeffrey E,
		Lucero E, Megrant A, Mutus J~Y, Neeley M, Neill C, O’Malley P~J~J, Quintana
		C, Roushan P, Sank D, Vainsencher A, Wenner J, White T~C, Solano E, Neven H
		and Martinis J~M 2016 {\em Nature\/} {\bf 534} 222
		
		\bibitem{Farhi:00}
		{Farhi} E, {Goldstone} J, {Gutmann} S and {Sipser} M 2000 {\em arXiv
			e-prints\/} quant-ph/0001106 (\textit{Preprint} \eprint{quant-ph/0001106})
		
		\bibitem{Farhi:01}
		Farhi E, Goldstone J, Gutmann S, Lapan J, Lundgren A and Preda D 2001 {\em
			Science\/} {\bf 292} 472--475
		
		\bibitem{Shor:96}
		Shor P 1996 Fault-tolerant quantum computation {\em Proceedings of 37th
			Conference on Foundations of Computer Science\/} pp 56--65
		
		\bibitem{Barends:14}
		Barends R, Kelly J, Megrant A, Veitia A, Sank D, Jeffrey E, White T~C, Mutus J,
		Fowler A~G, Campbell B, Chen Y, Chen Z, Chiaro B, Dunsworth A, Neill C,
		O’Malley P, Roushan P, Vainsencher A, Wenner J, Korotkov A~N, Cleland A~N
		and Martinis J~M 2014 {\em Nature\/} {\bf 508} 500--503
		
		\bibitem{Terhal:15}
		Terhal B~M 2015 {\em Rev. Mod. Phys.\/} {\bf 87}(2) 307--346
		
		\bibitem{Nielsen:Book}
		Nielsen M~A and Chuang I~L 2011 {\em Quantum Computation and Quantum
			Information: 10th Anniversary Edition\/} 10th ed (New York, NY, USA:
		Cambridge University Press) ISBN 1107002176, 9781107002173
		
		\bibitem{McClean_2016}
		McClean J~R, Romero J, Babbush R and Aspuru-Guzik A 2016 {\em New Journal of
			Physics\/} {\bf 18} 023023
		
		\bibitem{Cerezo:21}
		{Cerezo} M, {Arrasmith} A, {Babbush} R, {Benjamin} S~C, {Endo} S, {Fujii} K,
		{McClean} J~R, {Mitarai} K, {Yuan} X, {Cincio} L and {Coles} P~J 2021 {\em
			Nature Reviews Physics\/} {\bf 3} 625--644
		
		\bibitem{farhi_quantum_2014}
		{Farhi} E, {Goldstone} J and {Gutmann} S 2014 {A Quantum Approximate
			Optimization Algorithm} (\textit{Preprint} \eprint{1411.4028})
		
		\bibitem{Peruzzo:14}
		Peruzzo A, McClean J, Shadbolt P, Yung M~H, Zhou X~Q, Love P~J, Aspuru-Guzik A
		and O’brien J~L 2014 {\em Nature communications\/} {\bf 5} 4213
		
		\bibitem{Zhou:20}
		Zhou L, Wang S~T, Choi S, Pichler H and Lukin M~D 2020 {\em Phys. Rev. X\/}
		{\bf 10}(2) 021067
		
		\bibitem{Bharti:22}
		Bharti K, Cervera-Lierta A, Kyaw T~H, Haug T, Alperin-Lea S, Anand A, Degroote
		M, Heimonen H, Kottmann J~S, Menke T, Mok W~K, Sim S, Kwek L~C and
		Aspuru-Guzik A 2022 {\em Rev. Mod. Phys.\/} {\bf 94}(1) 015004
		
		\bibitem{Larkin:22}
		Larkin J, Jonsson M, Justice D and Guerreschi G~G 2022 {\em Quantum Science and
			Technology\/} {\bf 7} 045014
		
		\bibitem{Zhou:23}
		Zhou Z, Du Y, Tian X and Tao D 2023 {\em Phys. Rev. Appl.\/} {\bf 19}(2) 024027
		
		\bibitem{Zhang:25}
		Zhang Z, Paredes R, Sundar B, Quiroga D, Kyrillidis A, Duenas-Osorio L, Pagano
		G and Hazzard K~R~A 2024 {\em Quantum Science and Technology\/} {\bf 10}
		015022
		
		\bibitem{Lin:22}
		Lin J, Zhang Z, Zhang J and Li X 2022 {\em Phys. Rev. A\/} {\bf 105}(6) 062455
		
		\bibitem{Ozaeta:22}
		Ozaeta A, van Dam W and McMahon P~L 2022 {\em Quantum Science and Technology\/}
		{\bf 7} 045036
		
		\bibitem{Tabares:23}
		Tabares C, Mu\~noz de~las Heras A, Tagliacozzo L, Porras D and
		Gonz\'alez-Tudela A 2023 {\em Phys. Rev. Lett.\/} {\bf 131}(7) 073602
		
		\bibitem{Tabares:25}
		Tabares C, Kokail C, Zoller P, Gonz\'alez-Cuadra D and Gonz\'alez-Tudela A 2025
		{\em PRX Quantum\/} {\bf 6}(3) 030356
		
		\bibitem{Tabares:26}
		{Tabares} C, {Wild} D~S, {Cirac} J~I, {Zoller} P, {Gonz{\'a}lez-Tudela} A and
		{Gonz{\'a}lez-Cuadra} D 2025 {\em arXiv e-prints\/} arXiv:2511.04434
		(\textit{Preprint} \eprint{2511.04434})
		
		\bibitem{Hu:25}
		{Hu} C~K, {Xie} G, {Poulsen} K, {Zhou} Y, {Chu} J, {Liu} C, {Zhou} R, {Yuan} H,
		{Shen} Y, {Liu} S, {Zinner} N~T, {Tan} D, {Santos} A~C and {Yu} D 2025 {\em
			Nature Communications\/} {\bf 16} 3289
		
		\bibitem{anschuetz_variational_2018}
		Anschuetz E~R, Olson J~P, Aspuru-Guzik A and Cao Y 2018 Variational {Quantum}
		{Factoring} arXiv:1808.08927 [quant-ph]
		
		\bibitem{hegade_digitized_2021}
		Hegade N~N, Paul K, Albarrán-Arriagada F, Chen X and Solano E 2021 {\em
			Physical Review A\/} {\bf 104} L050403
		
		\bibitem{Hegade:22}
		Hegade N~N, Chen X and Solano E 2022 {\em Phys. Rev. Res.\/} {\bf 4}(4) L042030
		
		\bibitem{Chermoshentsev:21}
		{Chermoshentsev} D~A, {Malyshev} A~O, {Esencan} M, {Tiunov} E~S, {Mendoza} D,
		{Aspuru-Guzik} A, {Fedorov} A~K and {Lvovsky} A~I 2021 {\em arXiv e-prints\/}
		arXiv:2106.13167
		
		\bibitem{Nagies:25}
		Nagies S, Geier K~T, Akram J, Bantounas D, Johanning M and Hauke P 2025 {\em
			Quantum Science and Technology\/} {\bf 10} 035008
		
		\bibitem{Babbush:13}
		Babbush R, O'Gorman B and Aspuru-Guzik A 2013 {\em Annalen der Physik\/} {\bf
			525} 877--888
		
		\bibitem{Agresti:25}
		Agresti I, Paul K, Schiansky P, Steiner S, Yin Z, Pentangelo C, Piacentini S,
		Crespi A, Ban Y, Ceccarelli F, Osellame R, Chen X and Walther P 2025 {\em
			Phys. Rev. Res.\/} {\bf 7}(4) 043021
		
		\bibitem{albash_adiabatic_2018}
		Albash T and Lidar D~A 2018 {\em Reviews of Modern Physics\/} {\bf 90} 015002
		ISSN 0034-6861, 1539-0756
		
		\bibitem{Harrigan:21}
		{Harrigan} M~P, {Sung} K~J, {Neeley} M, {Satzinger} K~J, {Arute} F, {Arya} K,
		{Atalaya} J, {Bardin} J~C, {Barends} R, {Boixo} S, {Broughton} M, {Buckley}
		B~B, {Buell} D~A, {Burkett} B, {Bushnell} N, {Chen} Y, {Chen} Z, {Ben
			Chiaro}, {Collins} R, {Courtney} W, {Demura} S, {Dunsworth} A, {Eppens} D,
		{Fowler} A, {Foxen} B, {Gidney} C, {Giustina} M, {Graff} R, {Habegger} S,
		{Ho} A, {Hong} S, {Huang} T, {Ioffe} L~B, {Isakov} S~V, {Jeffrey} E, {Jiang}
		Z, {Jones} C, {Kafri} D, {Kechedzhi} K, {Kelly} J, {Kim} S, {Klimov} P~V,
		{Korotkov} A~N, {Kostritsa} F, {Landhuis} D, {Laptev} P, {Lindmark} M, {Leib}
		M, {Martin} O, {Martinis} J~M, {McClean} J~R, {McEwen} M, {Megrant} A, {Mi}
		X, {Mohseni} M, {Mruczkiewicz} W, {Mutus} J, {Naaman} O, {Neill} C, {Neukart}
		F, {Niu} M~Y, {O'Brien} T~E, {O'Gorman} B, {Ostby} E, {Petukhov} A,
		{Putterman} H, {Quintana} C, {Roushan} P, {Rubin} N~C, {Sank} D, {Skolik} A,
		{Smelyanskiy} V, {Strain} D, {Streif} M, {Szalay} M, {Vainsencher} A, {White}
		T, {Yao} Z~J, {Yeh} P, {Zalcman} A, {Zhou} L, {Neven} H, {Bacon} D, {Lucero}
		E, {Farhi} E and {Babbush} R 2021 {\em Nature Physics\/} {\bf 17} 332--336
		
		\bibitem{repo}
		See complete data and analysis in Python notebooks for this work from Felip
		Pellicer's \href{https://zenodo.org/records/18414560}{repository}.
		
		\bibitem{Pablo:26}
		D\'{\i}ez-Valle P, G\'omez-Ruiz F~J, Porras D and Garc\'{\i}a-Ripoll J~J 2026
		{\em Phys. Rev. Res.\/} {\bf 8}(1) 013211
		\urlprefix\url{https://link.aps.org/doi/10.1103/hxv2-sbr7}
		
		\bibitem{Santos:25}
		Santos A~C 2025 {\em Phys. Rev. A\/} {\bf 111}(2) 022618
		
		\bibitem{Lyngfelt:25}
		Lyngfelt I and Garc\'{\i}a-\'Alvarez L 2025 {\em Phys. Rev. A\/} {\bf 111}(2)
		022418
		
		\bibitem{McClean:18}
		{McClean} J~R, {Boixo} S, {Smelyanskiy} V~N, {Babbush} R and {Neven} H 2018
		{\em Nature Communications\/} {\bf 9} 4812
		
		\bibitem{Arrasmith:22}
		Arrasmith A, Holmes Z, Cerezo M and Coles P~J 2022 {\em Quantum Science and
			Technology\/} {\bf 7} 045015
		
		\bibitem{sarandy_consistency_2004}
		Sarandy M~S, Wu L~A and Lidar D~A 2004 {\em Quantum Information Processing\/}
		{\bf 3} 331--349
		
		\bibitem{Amin:09}
		Amin M~H~S 2009 {\em Phys. Rev. Lett.\/} {\bf 102}(22) 220401
		
		\bibitem{Stein:23}
		Stein J, Chamanian F, Zorn M, N\"{u}\ss{}lein J, Zielinski S, K\"{o}lle M and
		Linnhoff-Popien C 2023 Evidence that pubo outperforms qubo when solving
		continuous optimization problems with the qaoa {\em Proceedings of the
			Companion Conference on Genetic and Evolutionary Computation\/} GECCO '23
		Companion (New York, NY, USA: Association for Computing Machinery) p
		2254–2262 ISBN 9798400701207
		
		\bibitem{Albash:2012}
		Albash T, Boixo S, Lidar D~A and Zanardi P 2012 {\em New Journal of Physics\/}
		{\bf 14} 123016
		
		\bibitem{Albash:15}
		{Albash} T and {Lidar} D~A 2015 {\em Phys. Rev. A\/} {\bf 91} 062320
		
	\end{thebibliography}
	%\bibliographystyle{iopart-num}
	
	\providecommand{\newblock}{}

	\appendix
	\RefA{
		\section{QAOA Numerical Simulation Settings}\label{Appendix:SimulationSettings}
	}
	%We implement both protocols by exact state-vector simulation, applying each QAOA layer as an explicit matrix multiplication. The variational parameters are trained incrementally, initializing the depth-$p$ circuit from the optimized depth-$(p-1)$ parameters and seeding each new layer with a fixed heuristic. The initial states, the seeding heuristic, and the angle-folding constraint that sets the $p=1$ parameters are detailed in \ref{Appendix:SimulationSettings}. The cost is minimized with the unbounded gradient-based BFGS optimizer, which outperforms both gradient-free methods and the bounded \mbox{L-BFGS-B} variant across all instances up to eight qubits. To ensure a fair comparison, the same best-performing configuration is applied to both the PUBO and QUBO formulations, so that the standard protocol is evaluated under its most favorable setup.
	
	\RefA{All simulations are performed by exact state-vector evolution implemented in NumPy. The implementation is thus noiseless, and the approach remains efficient at the sizes considered ($n \le 8$ qubits). The complete implementation and analysis scripts are available in the repository~\cite{repo}.}
	
	\RefA{\paragraph{Initial states.} The initial state is taken to be an eigenstate of the mixer Hamiltonian $\hat{H}_\mathrm{M}$ occupying the same spectral region as the target solution under the problem Hamiltonian $\hat{H}_\mathrm{P}$. For the standard protocol the solution is the ground state, and we initialize in the mixer ground state $\ket{\Psi_0} = \ket{+}^{\otimes n}$. For the linearized protocol the solution lies in the center of the spectrum, and we initialize in the centred mixer eigenstate $\ket{\Psi_0} = \ket{+-+-\cdots}$. For an odd number of qubits this state does not belong to the kernel subspace of the Hamiltonian, but the variational optimization is robust to such a the small deviation of the initial state from the target energy subspace---as demonstrated through the high performance of the protocols shown in Fig.~\ref{Fig:Scaling}.}
	
	\RefA{\paragraph{Incremental training and parameter initialization.} The variational
		parameters are optimized layer by layer. The $p$-th layer of the circuit is initialized
		from the converged parameters obtained in the $(p-1)$-th layer, with the new layer initialized
		separately. The parameters $(\gamma_1, \beta_1)$ for the first layer are chosen from a scan of
		the $p=1$ cost landscape. The accessible range of $\gamma_1$ is bounded by angle
		folding, as the phase $e^{-i\gamma E_n}$ completes a full wrap
		once the largest eigenvalue satisfies $\gamma\, E_{\max} = 2\pi$, so we restrict
		$\gamma_1 \lesssim 2\pi/E_{\max}$. Because the PUBO spectrum spans a much larger
		range than the linearized one, this constraint forces markedly smaller
		$\gamma_1$ for the standard protocol than for the linear protocols. For each
		added layer we initialize $\gamma_{p+1} \leftarrow \tilde{\gamma}_p$ and
		$\beta_{p+1} \leftarrow 0$, where $\tilde{\gamma}_p$ is the optimized value of
		the previous final layer. 
		
		We found this heuristic to outperform the linear
		interpolation of the optimized-parameter trajectory commonly used for MaxCut-
		and QUBO-type problems~\cite{Zhou:20}, which assumes a smooth, adiabatic-like
		evolution of the optimal parameters and frequently traps the linearized
		protocol in local minima.}
	
	\RefA{\paragraph{Classical optimizer.} We optimize the variational parameters using
		gradient-based methods, comparing the bounded \mbox{L-BFGS-B} and the unbounded BFGS.
		Across all instances up to eight qubits, BFGS outperformed \mbox{L-BFGS-B}, which we attribute to its unbounded parameter range allowing it to escape minima that trap the bounded optimizer. The same optimizer is considered to both the PUBO and QUBO protocols, guaranteeing a fair comparison between the formulations.
	}
	
	\section{Table of factorized numbers}\label{Appendix:Table}
	
	The details about all numbers factorized in our work are shown in Table~\ref{Table_factor} below.
	
	\begin{table}[h]
		\centering
		\caption{\label{Table_factor}Overview of all semiprime instances considered in this work, including the number of qubits $n$ required for the QAOA, the values of $p$, $p^{\prime}$, $q$ and $q^{\prime}$, the binary number representation for $p'$ and $q'$, as well as the expected outcome $\ket{\psi_{\mathrm{out}}}$ in the computational basis.}
		\footnotesize
		\begin{tabular}{@{}ccccccccc@{}}
			\br
			$N$ & $n$ & $p$ ($p'$) & $q$ ($q'$) & $n_p$ & $n_q$ & $p_\mathrm{bitstring}$ & $q_\mathrm{bitstring}$ & $\ket{\psi_{\mathrm{out}}}$ \\
			\mr
			15  & 3          & 3 (1)      & 5 (2)      & 1     & 2     & 1                      & 10                     & $\ket{101}$                   \\
			21  & 3          & 3 (1)      & 7 (3)      & 1     & 2     & 1                      & 11                     & $\ket{111}$                   \\
			25  & 4          & 5 (2)      & 5 (2)      & 2     & 2     & 10                     & 10                     & $\ket{0101}$                  \\
			35  & 5          & 5 (2)      & 7 (3)      & 2     & 3     & 10                     & 011                    & $\ket{01110}$, $\ket{11010}$          \\
			39  & 5          & 3 (1)      & 13 (6)     & 2     & 3     & 01                     & 110                    & $\ket{10011}$                 \\
			51  & 6          & 3 (1)      & 17 (8)     & 2     & 4     & 01                     & 1000                   & $\ket{100001}$                \\
			77  & 6          & 7 (3)      & 11 (5)     & 2     & 4     & 11                     & 0101                   & $\ket{111010}$                \\
			87  & 7          & 3 (1)      & 29 (14)    & 3     & 4     & 001                    & 1110                   & $\ket{1000111}$               \\
			95  & 7          & 5 (2)      & 19 (9)     & 3     & 4     & 010                    & 1001                   & $\ket{0101001}$               \\
			115 & 8          & 5 (2)      & 23 (11)    & 3     & 5     & 010                    & 01011                  & $\ket{01011010}$              \\
			119 & 8          & 7 (3)      & 17 (8)     & 3     & 5     & 011                    & 01000                  & $\ket{11000010}$              \\
			143 & 8          & 11 (5)     & 13 (6)     & 3     & 5     & 101                    & 00110                  & $\ket{10101100}$, $\ket{01110100}$    \\
			\br
		\end{tabular}
	\end{table}
\end{document}